# Mapping the structural transitions controlled by the trilinear coupling in $Ca_{3-x}Sr_xTi_2O_7$

Marie Kratochvilova[1,2,3], Fei-Ting Huang[4], Maria-Teresa Fernandez Diaz[5], Milan Klicpera[3], Sarah J. Day[6], Stephen P. Thompson[6], Yoon-Seok Oh[7], Bin Gao[4], Sang-Wook Cheong[4], and Je-Geun Park[1,2]

[1]*Center for Correlated Electron Systems, Institute for Basic Science, Seoul 08826, Korea*
[2]*Department of Physics and Astronomy, Seoul National University, Seoul 08826, Korea*
[3]*Charles University, Faculty of Mathematics and Physics, Department of Condensed Matter Physics, Ke Karlovu 5, 121 16 Prague 2, Czech Republic*
[4]*Rutgers Center for Emergent Materials and Department of Physics and Astronomy, Rutgers University, Piscataway New Jersey 08854, USA*
[5]*Institut Laue Langevin, 71 Avenue des Martyrs, CS 20156, 38042 Grenoble Cedex 9, France*
[6]*Diamond Light Source, Harwell Science and Innovation Campus, Didcot OX11 0DE, Oxon, England*
[7]*Department of Physics, Ulsan National Institute of Science and Technology (UNIST), Ulsan 689-798, Korea*

We present the results of the high-temperature neutron and x-ray diffraction experiments on the $Ca_{3-x}Sr_xTi_2O_7$ ($x$ = 0.5, 0.8, 0.85, 0.9) compounds. The ferro- to paraelectric transition in these hybrid improper ferroelectric materials arises from the so-called trilinear coupling. Depending on the Strontium content, various structures and phase transitions, different from theoretical predictions, emerge. The *in-situ* x-ray powder diffraction indicates a direct ferro- to paraelectric transition between the orthorhombic $A2_1am$ and tetragonal undistorted $I4/mmm$ phase for $x \leq 0.6$. We identified a reduction in the trilinear coupling robustness by increasing the Sr-doping level to lead to the emergence of the intermediate tetragonal $P4_2/mnm$ phase and the gradual suppression of the orthorhombic phase. The observed character of the structure transitions and the $Ca_{3-x}Sr_xTi_2O_7$ phase diagram are discussed in the framework of theoretical models of other related hybrid improper ferroelectric systems.

## I. INTRODUCTION

Proper ferroelectrics are characterized by polarization which is the primary order parameter originating from zone-center polar lattice instability. On the other hand, in hybrid improper ferroelectrics (HIF), polarization is part of a more complex lattice distortion arising from a combination of two nonpolar lattice modes with different symmetries, the so-called trilinear coupling [1,2]. In perovskites, these modes correspond to oxygen polyhedral distortions, commonly rotations or tilts which can, in addition, induce magnetoelectricity and weak ferromagnetism. Such an electric field-controllable mechanism can be extremely attractive for device applications if operated at room temperature and hereby attracts enormous attention [3]. Moreover, it has been shown for $Ca_{3-x}Sr_xTi_2O_7$ crystals recently, that there is a unique domain topology associated with various types of oxygen octahedral distortions having direct impact on ferroelectric properties [4,5,6].

The family of HIF compounds reveals an amazingly rich set of structure transitions, as was demonstrated on Ca$_{3-x}$Sr$_x$Ti$_2$O$_7$ [4], Sr$_3$Zr$_2$O$_7$ [7], and Ca$_3$Mn$_2$O$_7$ [8], recently. These materials belong to the Ruddlesden-Popper phases described by the general formula A$_{n+1}$B$_n$O$_{3n+1}$ ($n = 2$). The structure consists of perovskite ABO$_3$ building blocks stacked along the [001]-axis with a rocksalt AO layer interspersing every two perovskite unit cells. In the ferroelectric (FE) state, A$_3$B$_2$O$_7$ forms the orthorhombic polar space group *A*2$_1$*am* which corresponds to the distorted tetragonal *I*4/*mmm* structure where the paraelectric (PE) phase is eventually established upon heating. How the orthorhombic structure transforms into the distortion-free aristotype phase upon heating has been the subject of many theoretical works [1,2,9]. Experimentally, it has been demonstrated that each of these compounds exhibits its own unique route (summarized in TABLE I.).

Improper hybrid ferroelectricity has been experimentally demonstrated for the first time in the orthorhombic Sr-doped Ca$_3$Ti$_2$O$_7$ [3]. This distortion includes the rotation of the BO$_6$ octahedra around the [001]-axis described by the in-phase rotation mode $X_2^+$ and the diagonal (with respect to the tetragonal-like basal plane) tilting mode $X_3^-$ around the [110]-axis, which are simultaneously coupled to the $\Gamma_5^-$ polar zone-center mode thereby creating the trilinear coupling [1]. By introducing the Sr-substitution one can tune the robustness of distortions as the paraelectric Sr$_3$Ti$_2$O$_7$ phase crystallizes in the *I*4/*mmm* space group. Theoretical works predicted the existence of an intermediate phase, either *Amam* or *Acam* [1,9]; however, the combined study of Ca$_{3-x}$Sr$_x$Ti$_2$O$_7$ provided a different picture [4] by discovering the intermediate paraelectric *P*4$_2$/*mnm* phase in Ca$_2$SrTi$_2$O$_7$ characterized by the $X_3^-$ [100] tilting mode oriented along the tetragonal-like directions. The impact of the discovered structure on physical properties has been studied experimentally [10] and also theoretically by a first-principles study recently [11]. The *A*2$_1$*am*-*P*4$_2$/*mnm* transition corresponds to the relaxation of the in-phase rotation $X_2^+$ [4]. Crossing the *P*4$_2$/*mnm*-*I*4/*mmm* border is related to the relaxation of the $X_3^-$ tilting mode.

Sr$_3$Zr$_2$O$_7$ undergoes a first-order FE-PE transition, where the paraelectric polymorph competes with the polar phase and emerges from a trilinear coupling of rotation and tilt modes interacting with an antipolar mode [7]. The structure transforms from *A*2$_1$*am* to *I*4/*mmm* via two intermediate paraelectric phases upon heating: First, the $X_1^-$ rotation mode is lost in the second-order *Pnab* → *Amam* transition and then the tilting mode $X_3^-$ is lost in the second-order *Amam* → *I*4/*mmm* transition. The *A*2$_1$*am* → *Pnab* structure transition reveals hysteresis corresponding to a first-order transition linked to the competition of the $X_1^-$ out-of-phase and the $X_2^+$ in-phase rotations.

In Ca$_3$Mn$_2$O$_7$, the high-temperature *I4/mmm* structure undergoes a transition to an intermediate PE orthorhombic *Acaa* structure upon cooling and then changes to *A2$_1$am*. [12]. As the temperature is further lowered, an antiferromagnetic order sets in at 115 K. The *Acaa* symmetry is generated by the out-of-phase rotation $X_1^-$ thus, the transition pathway from *Acaa* to *A2$_1$am* results in the competition of the $X_1^-$ out-of-phase and $X_2^+$ in-phase rotations, thereby related to peculiar physical phenomena [8]. It is worth noting that the Mn substitution on the Ti-site of Ca$_3$Ti$_2$O$_7$ has also been studied both experimentally [13,14] and theoretically [15]; for the Ti-rich side, *I4/mmm* goes directly to polar *A2$_1$am* symmetry upon cooling while the structural transition processes via the *Acaa* symmetry for the Mn-rich side [13].

In this work, we present the temperature evolution of structure phase transitions revealed by varying the Strontium concentration in $Ca_{3-x}Sr_xTi_2O_7$ ($0.5 \leq x \leq 0.9$) using *in-situ* high-temperature x-ray diffraction. We show that on the Ca-rich side the structure transition processes directly from *I4/mmm* to polar *A2$_1$am* ($x < 0.8$) while the intermediate phase *P4$_2$/mnm* is preferred in the transition pathway for the Sr-rich compositions. The structure-Sr concentration phase diagram is constructed and discussed within the framework of related HIF compounds.

TABLE I. The temperature evolution of the crystal structures with corresponding irreducible representations in various HIF systems [4,7,8].

| | *T* increases → | | | |
|---|---|---|---|---|
| $Ca_{3-x}Sr_xTi_2O_7$ | $X_2^+ \oplus X_3^-$ | $X_3^-$ | - | |
| | *A2$_1$am* | *P4$_2$/mnm* | *I4/mmm* | |
| $Sr_3Zr_2O_7$ | $X_2^+ \oplus X_3^-$ | $X_1^- \oplus X_3^-$ | $X_3^-$ | - |
| | *A2$_1$am* | *Pnab* | *Amam* | *I4/mmm* |
| $Ca_3Mn_2O_7$ | $X_2^+ \oplus X_3^-$ | $X_1^-$ | - | |
| | *A2$_1$am* | *Acaa* | *I4/mmm* | |

## II. EXPERIMENTAL DETAILS

The polycrystalline $Ca_{3-x}Sr_xTi_2O_7$ samples ($x = 0.5, 0.8, 0.9$) were prepared using a solid-state reaction method as described in Ref. [4]. The $Ca_{2.15}Sr_{0.85}Ti_2O_7$ single crystals were grown by an optical floating zone method [3]. The samples with $x = 0.5, 0.8$ and $0.9$ were characterized using the high-resolution x-ray diffractometer (Bruker XRD D8 Discover with Cu K$\alpha$ source) equipped with the high-temperature chamber HTK 1200N. Additionally, the sample with $x = 0.5$ was measured on the high resolution synchrotron powder diffraction beamline I11 at Diamond Light Source using the multi-analyzer crystal detector stages and 15 keV x-rays calibrated against NIST SRM 640c silicon powder. Samples were loaded in quartz capillaries and variable temperature heating provided by a cyberstar hot air blower [16]. The neutron diffraction of pulverized single crystals with $x = 0.85$ was measured on the D2B high-resolution powder diffractometer (ILL). The single-crystal diffraction on the $x = 0.85$ sample from a different batch was measured using the hot neutron four-circle D9 diffractometer (ILL) equipped with a four-cycle furnace. The sets of D2B and D9 data can be found in Ref. [10.5291/ILL-DATA.5-15-610]. The diffraction data was analyzed using the FullProf [17] software. The differential scanning calorimetry (DSC) was measured using SETSYS Evolution 24 instrument (SETARAM) in He atmosphere. The heating/cooling rate was 10 K/min and the transition temperatures were determined by the onset of the observed peaks.

Powder neutron diffraction study of a sample with $x = 0.85$ revealed the decay of the (237) and (033) reflections at ~ 550 °C upon increasing temperature as shown in Fig. 1 (a,b). The single-crystal neutron diffraction of a sample from a different batch observed the decay of the (033) Bragg peak at slightly higher temperature ~ 580 °C which might be caused by a negligible variation of the Strontium concentration.

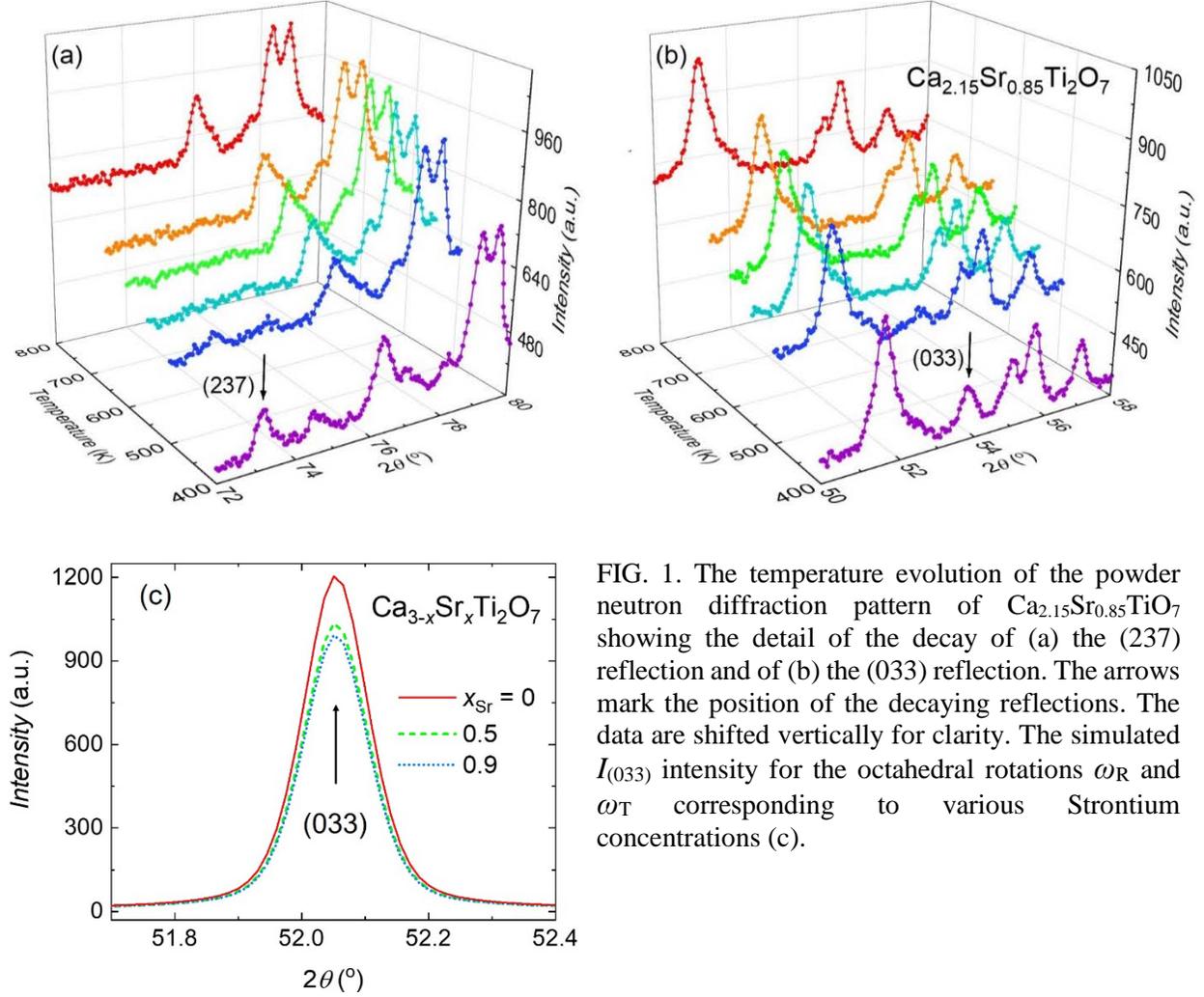

FIG. 1. The temperature evolution of the powder neutron diffraction pattern of $Ca_{2.15}Sr_{0.85}TiO_7$ showing the detail of the decay of (a) the (237) reflection and of (b) the (033) reflection. The arrows mark the position of the decaying reflections. The data are shifted vertically for clarity. The simulated $I_{(033)}$ intensity for the octahedral rotations $\omega_R$ and $\omega_T$ corresponding to various Strontium concentrations (c).

## III. RESULTS AND DISCUSSIONS

As the (237) reflection has a lower intensity and higher $2\theta$ in the x-ray diffraction data compared to the (033) reflection, the latter was chosen to identify transitions across various structure phases. The integrated intensities $I_{(033)}$ of the superlattice peaks are expected to scale as the square of the order parameter $Q$ for these phase transitions, where $Q$ is given by the rotation and tilting angles $\omega_R$ and $\omega_T$ of the oxygen octahedra, respectively. The angles are given by the Strontium concentration as was shown by the synchrotron x-ray diffraction experiment [4]. Simulation of the (033) reflection integrated intensity for different octahedral rotations (tilts) which corresponds to the increasing Strontium concentration is shown in Fig. 1 (c). It clearly shows the decrease of intensity by ~15 % for the sample with $x = 0.8$ and ~18 % for sample with $x = 0.9$ from the undoped $Ca_3Ti_2O_7$. The values of $I_{(033)}$ were obtained by fitting of the peak profile using the Lorentz function. The (033) reflection of the $x = 0.8, 0.85$, and $0.9$ samples was scanned in the $2\theta$ region from 51.1° to 53.1° upon cooling using the Bruker diffractometer. In case of the $x = 0.5$ sample, the reflection was tracked in the $2\theta$ region from 16.0° to 16.5° upon warming using the I11 instrument. The resulting scans upon decreasing temperature are summarized in Fig. 2. In all four compounds, a gradual decrease of the (033) intensity with increasing temperature is revealed. The free energy of the HIF superlattice system in terms of Ginzburg-Landau (GL) theory [18,19] is given by

$$G = A_1(T - T_C)\omega^2 + A_2\omega^4 + A_3\omega^6 + C_3 P \omega_R \omega_T + C_4 P^2 - PE, \qquad (1)$$

where $\omega = (\frac{1}{\sqrt{2}}\omega_R, \frac{1}{\sqrt{2}}\omega_R, \omega_T)$, $E$ is the external electric field, and $P$ is the electrical polarization of $Ca_3Ti_2O_7$. The quadratic term and the biquadratic terms coupling the rotations and polarization stated in Eq. (1) [19] are omitted with respect to the trilinear term for temperatures close to $T_C$. Assuming the first-order nature of the phase transition and the collaborative trilinear interaction, the coefficients $A_1$, $A_3$, and $C_4$ are expected to be positive while $A_2$ and $C_3$ negative, respectively. The equilibrium solution providing non-negative $\omega^2$ obtained by minimizing the free energy is

$$\omega^2 = \frac{1}{6A_3}(-2A_2 + (4A_2^2 - 12A_3[A_1(T - T_C) + C_3'P])^{1/2}), \qquad (2)$$

where the term $C_3'$ is defined analogically to Ref. [4], providing a first-order transition for $A_2 < 0$, tricritical transition for $A_2 = 0$, and a continuous transition for $A_2 > 0$. With respect to the $I_{(033)}(T)$ error bars, qualitative results. i. e. the signs are given for the fitting coefficients except the transition temperatures.

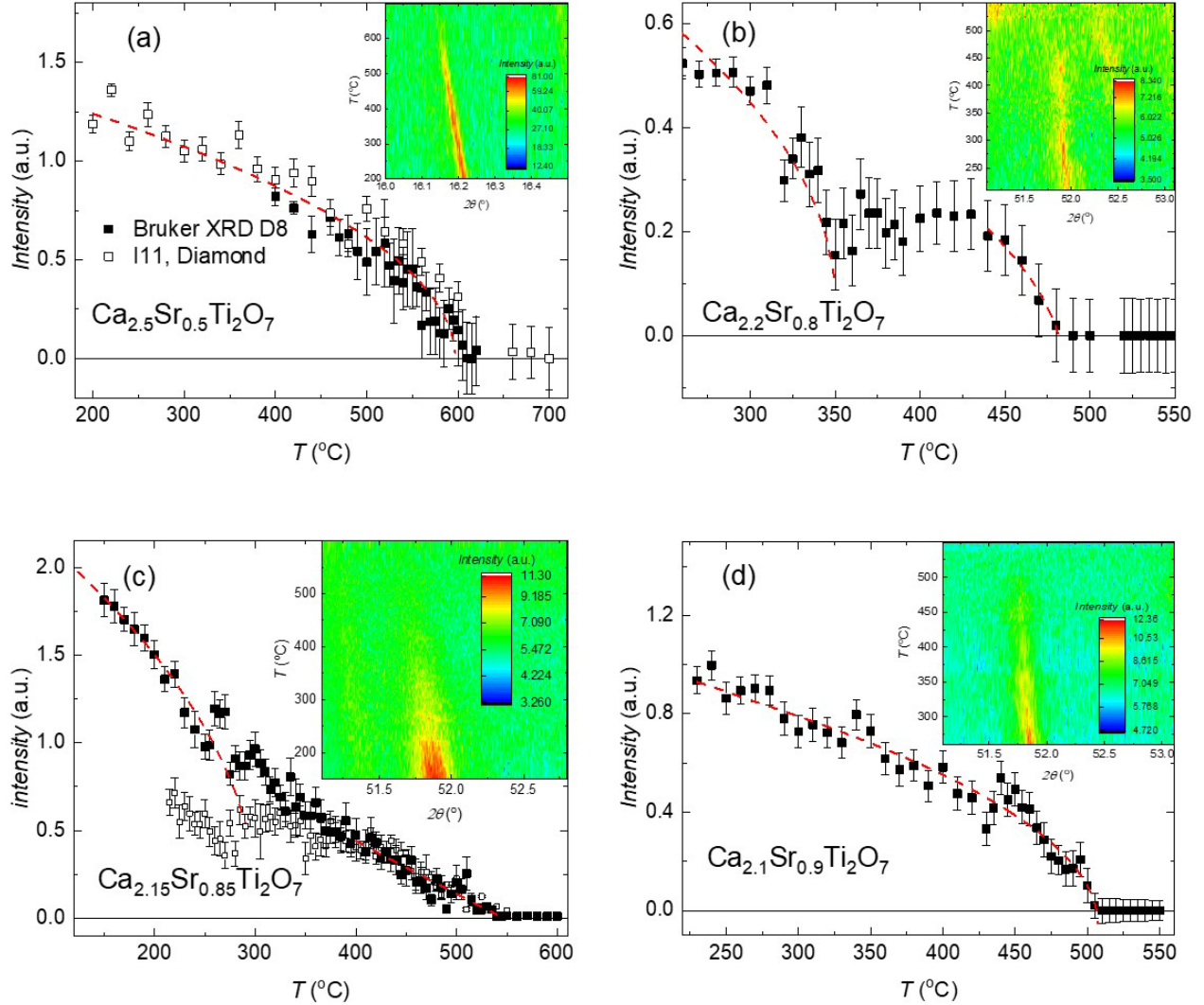

FIG. 2. The temperature dependence of the (033) reflection intensity of (a) $Ca_{2.5}Sr_{0.5}Ti_2O_7$, (b) $Ca_{2.2}Sr_{0.8}Ti_2O_7$, (c) $Ca_{2.15}Sr_{0.85}Ti_2O_7$, and (d) $Ca_{2.1}Sr_{0.9}Ti_2O_7$. The red dashed line is the GL fit. The aspects of the fitting procedure and the fitting parameters are discussed in the main text. The insets show temperature-intensity contour plots from the *in-situ* x–ray diffraction measured at I11 (a), and measured using the Bruker diffractometer (b-d). The data were collected while warming, except the data in (c) which were measured both upon cooling and warming.

In the case of $Ca_{2.5}Sr_{0.5}Ti_2O_7$, the intensity decreases monotonically down to background values at $T_C$ using data combined from the I11 and Bruker measurements as shown in Fig. 2 (a). The structure of the room-temperature phase corresponds to the $A2_1am$ space group while the phase above $T_C$ can be described by the $I4/mmm$ space group. The fit of the data for the $Ca_{2.5}Sr_{0.5}Ti_2O_7$ sample offer $T_C = 605(15)°C$, positive $A_1$ and $A_3$ parameters, negative $C'_3$ coefficient in line with the trilinear coupling, and the $A_2$ coefficient approaching zero values within the large error bars proposing the composition with $x_{Sr} = 0.5$ is very close to the tricritical point.

In Fig. 2 (b) and (c), we observe a different situation. Upon cooling, the temperature dependence of the (033) reflection intensity is characterized by a step-like structure emerging at

$T_{C1} = 360(25)$°C in $Ca_{2.2}Sr_{0.8}Ti_2O_7$ and $T_{C1} = 280(30)$°C in $Ca_{2.15}Sr_{0.85}Ti_2O_7$. This remarkable change of slope signifies a transition from the room-temperature $A2_1am$ space group to the tetragonal phase described by the $P4_2/mnm$ symmetry. The corresponding coefficients $A_1$ and $A_3$ are positive, while $C_3'$ and $A_2$ are negative for both systems, pointing to the first-order nature of the $A2_1am$-$P4_2/mnm$ structural transition as expected from the group-subgroup relations. The (033) intensity decays completely at $T_{C2} = 480(15)$°C in $Ca_{2.2}Sr_{0.8}Ti_2O_7$ ($T_{C2} = 550(18)$°C in $Ca_{2.15}Sr_{0.85}Ti_2O_7$) when the high-temperature $I4/mmm$ structure is established. The fitting parameter $A_2$ is clearly positive for $Ca_{2.2}Sr_{0.8}Ti_2O_7$ and $Ca_{2.15}Sr_{0.85}Ti_2O_7$, respectively, revealing a continuous $P4_2/mnm$-$I4/mmm$ transition in agreement with the Landau theory. To verify the order of the transitions obtained by the fits, we have measured the $Ca_{2.15}Sr_{0.85}Ti_2O_7$ sample both upon cooling and warming. A clear declination from the warming curve signifying hysteretic behavior and hence a discontinuous transition is observed below ~ 350°C, although lower temperatures could not be measured due to the gradual degradation of the sample in the furnace.

The intensity evolution in $Ca_{2.1}Sr_{0.9}Ti_2O_7$ is different from the previous two samples, as shown in Fig. 2 (d). Here, the smooth decay of the (033) intensity suggests a second-order, single structure transition between the two tetragonal phases $P4_2/mnm$ and $I4/mmm$ at $T_{C2} = 510(10)$°C with the positive $A_2$ parameter.

We further note that the DSC experiment on $Ca_{2.15}Sr_{0.85}Ti_2O_7$ does not show any anomaly up to 1300 °C (see Fig. 3(b)) which might be caused by the structure disorder increasing for higher Strontium concentration, smearing out the discontinuous character of the transition at $T_{C1}$ ~ 280°C. On the other hand, the heating and cooling curves obtained on $Ca_3Ti_2O_7$ (inset of Fig. 3(b)) still show a transition at $T_C$ ~ 760 °C with hysteresis $\Delta T$ ~ 10 °C which agrees well with the DSC data presented in Ref. [20].

The data are summarized in the temperature-concentration phase diagram shown in Fig. 3(a). It can be separated into three regions O, T', and T defined by the $A2_1am$, $P4_2/mnm$, and $I4/mmm$ space groups. Clearly, the Ca-rich side reveals a direct transition from the orthorhombic to the undistorted tetragonal structure, while a narrow region of the intermediate phase opens in the Sr concentration range $0.6 < x < 0.8$ and broadens with increasing $x$. According to Ref. [4], the undistorted $I4/mmm$ phase is finally stabilized for $x > 1$. The phase diagram includes data obtained from the *in situ* x-ray diffraction, *ex-situ* observation of twin-change under polarized optical microscopy (POM) after various heat treatments, *in situ* transmission electron microscopy (TEM) of $Ca_{2.1}Sr_{0.9}Ti_2O_7$ [4], and DSC (this work and [10]).

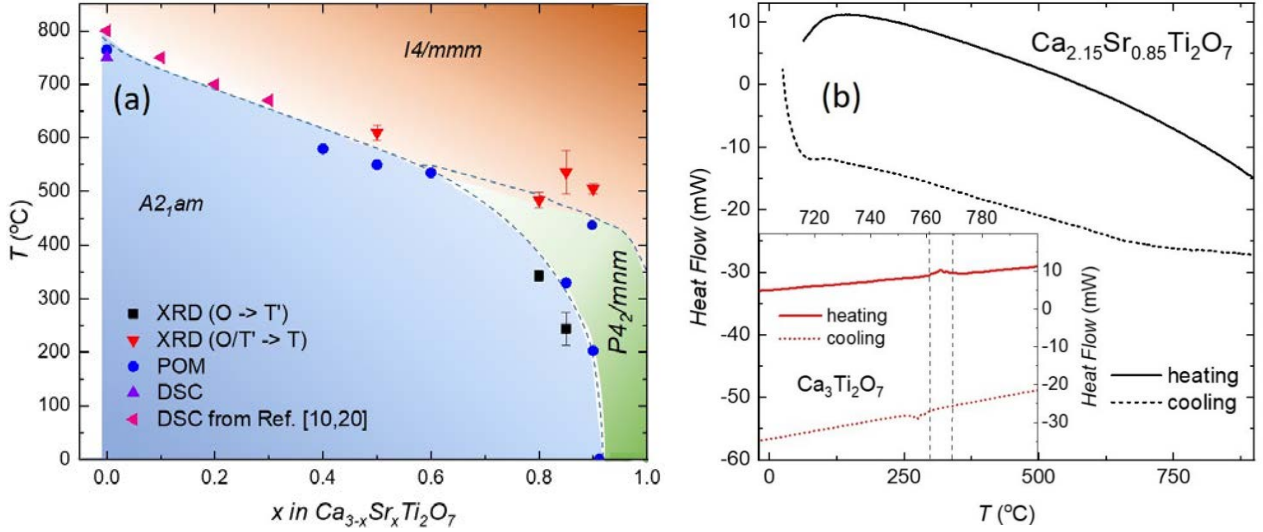

FIG. 3. (a) The temperature-Sr concentration phase diagram of the $Ca_{3-x}Sr_xTi_2O_7$ system. The squares mark the O→T' transition and the red triangles mark the O/T'→T transition from the x-ray diffraction. The other triangles mark the O→T transition from DSC (this work and Ref. [10,20]). The circles represent data ($x < 0.9$) from the ex-situ observation of twin-change under POM. The data point for $x = 0.9$ marked by the circle was obtained from in-situ TEM and the error bar is within the symbol. The solid lines are guides to the eye. (b) DSC curves of $Ca_{2.15}Sr_{0.85}Ti_2O_7$ measured up to 1300 °C. The inset shows the detail view of $Ca_3Ti_2O_7$ with the transition temperature ~ 760 °C marked by vertical lines. The solid and dotted lines represent the heating and cooling processes, respectively.

Turning our attention to the character of the structure transitions in HIF we observe an agreement between our and previously published data: a first-order phase transition has been observed on the DSC curves in $Ca_3Ti_2O_7$ [20] and $Ca_{3-x}Sr_xTi_2O_7$ for $x < 0.5$ [10], corresponding to the O→T crossover in line with the group-subgroup relations. Similarly, a discontinuous change of the tilting order parameter when crossing the O→T' boundary was reported in the Sr-rich composition with $x = 0.9$ [4,5]. The character of the transitions can be revealed also by the domain topology which has been heavily discussed recently [4,5,6]. The formation of $Z_2 \times Z_2$ domains with $Z_4$ vortices is associated with a second–order transition in the Sr-rich compounds. On the other hand, the phases with $x < 0.95$ form the $Z_4 \times Z_2$ domains with $Z_3$ vortices which are related to discontinuous transitions [5]. Our data show a discontinuous transition in case of the $A2_1am$-$P4_2/mnm$ paths when the trilinear term in the GL fit is considered, while a second-order transition for the $P4_2/mnm$-$I4/mmm$ crossover is observed in all measured compositions. Interestingly, the GL fit of the $A2_1am$-$I4/mmm$ transition in $Ca_{2.5}Sr_{0.5}Ti_2O_7$ suggests that this concentration might be on the verge of the tricritical point as indicated in the phase diagram.

In a loose analogy with the mean-field calculations using a simple microscopic Hamiltonian applied on the Aurivillius compounds, we can find a close resemblance with the $Ca_{3-x}Sr_xTi_2O_7$ phase diagram shown in Figure 2 [21] and the order of the structure transitions. The temperature evolution of the order parameters $\phi_1$ and $\phi_3$ (Figure 1(b), Ref. [21]) corresponds very well to the temperature evolution of the (033) intensity in $Ca_{2.15}Sr_{0.85}Ti_2O_7$ and $Ca_{2.2}Sr_{0.8}Ti_2O_7$. Depending on the coupling strength $\gamma$ (Strontium concentration in our case), the character of the O→T' and T'→T transitions can be either second- or first-order. Increasing $\gamma$ leads to a single first-order transition (Figure 1(c), Ref. [21]) which, in this simplified view, corresponds to the $Ca_{2.5}Sr_{0.5}Ti_2O_7$ data. In further analogy with these schemes, we should expect the T'→T transition to reveal a second-order character in $Ca_{2.1}Sr_{0.9}Ti_2O_7$, and both types of transitions can emerge in the intermediate compositions. Such a scenario implies the existence of a tricritical point, where a first-order transition changes into a second-order one. It has been suggested that the density of domain walls is expected to be smaller for a tricritical phase transition, compared to a second-order one [22]. The tricritical behavior near the $x_{Sr} = 0.5$ concentration demands experiments focused on the region of the O, T', and T phase coexistence investigating a variety of Strontium compositions.

## IV. CONCLUSIONS

Using *in-situ* high-temperature x-ray diffraction and other techniques, we have studied the structural transitions in the $Ca_{3-x}Sr_xTi_2O_7$ HIF compounds tuning the trilinear coupling strength by varying the Sr-doping level. Consistently with previous studies, the Ca-rich compositions reveal a single direct transition from the orthorhombic ferroelectric *A*2$_1$*am* phase to the tetragonal paraelectric *I*4/*mmm* phase. For $x \geq 0.8$, an intermediate paraelectric *P*4$_2$/*mnm* phase emerges wedged in between the room-temperature and the high-temperature phase as revealed by two successive transitions. Above $x \sim 0.9$, we observe a single phase transition from *P*4$_2$/*mnm* to *I*4/*mmm* at high temperatures. The temperature dependence of the $I_{(033)}$ intensities and the phase diagram corresponds qualitatively very well to the calculated diagrams of the Aurivillius phases. Using the GL fit with the trilinear term included we show that the *A*2$_1$*am*-*P*4$_2$/*mnm* structural transition across the series is first-order-like, while the *P*4$_2$/*mnm*-*I*4/*mmm* crossover reveals a continuous character, consistently with previous experiments. Moreover, the data suggest the presence of the tricritical behavior for the $x_{Sr} = 0.5$ concentration. We have also shown that the simple x-ray diffraction method can be used to identify the order of the transitions using the modified GL fit for the HIF systems. To clarify the related domain configuration and the properties of the domain walls in the vicinity of the critical point where the three phases merge, the next step is to prepare high-quality samples from the concentration range $0.6 < x < 0.8$ and explore this part of the $Ca_{3-x}Sr_xTi_2O_7$ phase diagram by further x-ray and neutron diffraction experiments.

## ACKNOWLEDGMENTS

This work was supported by the Institute for Basic Science (IBS) in Korea (IBS-R009-G1). We acknowledge Institute Laue-Langevin (ILL), Grenoble, France, for the allocation of time on D2B and D9 diffractometers. The work was supported within the program of Large Infrastructures for Research, Experimental Development and Innovation (project No. LM2015050) and project LTT17019 financed by the Ministry of Education, Youth and Sports, Czech Republic. We acknowledge the support and beam time award no. "EE16074" at Diamond Light Source in providing synchrotron research facilities used in this work. The DSC experiments were performed in MGML (https://mgml.eu/) through the program of Czech Research Infrastructures (LM2011025). The work at Rutgers University was supported by the DOE under Grant No. DOE: DE-FG02-07ER46382.